\documentclass[twocolumn,pra,superscriptaddress,amsmath,amssymb,mathrsfs,hyperref]{revtex4-2}
\usepackage{tabularx}
\usepackage{graphicx}
\usepackage{braket}
\usepackage[utf8]{inputenc}
\usepackage[normalem]{ulem}

\usepackage{dcolumn}
\usepackage{bm}
\usepackage{epsfig}
\usepackage{rotating}
\usepackage{color}
\usepackage{orcidlink}
\usepackage
{hyperref}

\hypersetup{
    colorlinks=true,
    linkcolor=blue,
    citecolor=blue,      
    urlcolor=blue
    }

\newcommand{\changed}[1]{{\color{black} #1}}
\newcommand{\rosi}[1]{{\color{black} #1}}

\begin{document}
\title{A linear rotor trapped and coupled to the vibrational modes of an ion crystal}

\author{Monika Leibscher\,
\orcidlink{0000-0001-6006-9067}}
\affiliation{Freie Universit\"{a}t Berlin, Fachbereich Physik and Dahlem Center for Complex Quantum Systems, Arnimallee 14, 14195 Berlin, Germany}
\author{Juan M. García-Garrido}
\affiliation{Departamento de Física Atómica, Molecular y Nuclear, Universidad de Granada, Granada, Spain}
\author{Konstantin Gaul}
\affiliation{Helmholtz Institut Mainz, 55099 Mainz, Germany}
\affiliation{GSI Helmholtzzentrum für Schwerionenforschung GmbH, 64291 Darmstadt, Germany}
\affiliation{QUANTUM, Institut für Physik, Johannes Gutenberg University, Staudingerweg 7, 55128 Mainz, Germany}
\author{Rosario González-Férez\,
\orcidlink{0000-0002-8871-116X}}
\affiliation{Departamento de Física Atómica, Molecular y Nuclear, Universidad de Granada, Granada, Spain}
\author{Ferdinand Schmidt-Kaler}
\affiliation{QUANTUM, Institut für Physik, Johannes Gutenberg University, Staudingerweg 7, 55128 Mainz, Germany}
\author{Christiane P. Koch\,
\orcidlink{0000-0001-6285-5766}}
\thanks{Corresponding author.}
\email{christiane.koch@fu-berlin.de}

\date{\today} 
 
\begin{abstract}
When molecular ions are trapped together with atomic ions in a Paul trap, their dipole moment couples the molecular rotation to the joint vibrational motion of the particles in the trap. To leverage this coupling for quantum control, a rotational transition in the molecule should be resonant with one of the ion crystal vibrational modes. Focusing on the example of singly charged thorium fluoride molecular ions, cotrapped with two ytterbium ions, we determine the conditions for resonant dipole-phonon coupling, fully accounting for the molecular hyperfine structure. We identify several choices for resonant coupling and discuss its detection using sideband-resolved laser spectroscopy and measurements of decoherence.
\end{abstract}

\maketitle

\section{Introduction}
Molecular ions have been proposed for applications in quantum information processing~\cite{wasielewski2020,zelevinsky2023}, for quantum simulation~\cite{KochRMP2019}, and to establish quantum error correction codes~\cite{AlbertPRX2020,asnaashari2023,
RaynalPRA2010,jain2024,XuPRA2024,vuillot2024}. Recent experimental progress in the field of cold trapped molecular ions \cite{Deiss2023,Sin22b} includes the preparation and coherent manipulation of quantum states of a single molecular ion \cite{ChouNature2017}, the non-destructive state detection via 
co-trapped ions \cite{WolfNature_2016,SinhalScience2020}, and quantum logic operations \cite{HudsonPRA2018,Ni2018,Campell_PRL_2020}.
The size and mass range of molecules that can be trapped and cooled in Paul traps is impressive, spanning from diatomics to particles as large as biomolecules \cite{OffenbergPRA2008,OffenbergJPhysB2009}.

Diatomic molecular ions in particular are currently attracting significant interest: High precision spectroscopy on polar diatomic molecular ions such as HD$^+$, HfF$^+$, or ThF$^+$ is becoming a laboratory platform for fundamental physics, allowing for tests of the standard model and physics beyond the standard model~\cite{Cairncross_2017,Ng2022,Cal2024,Schiller2024,Fan_PRA_2025}. 
Multiply charge thorium fluoride ions can be produced \cite{Stricker_2025}, with 
singly charged ThF$^+$ attracting interest as a promising candidate for the search for the electron’s electric dipole moment~\cite{Meyer_2008, Gresh_2016, Zhou_2019, Ng2022}, eventually combined with with quantum locic spectroscopy~\cite{Zhou24} and time-reversal violation~\cite{Denis_2015}. Moreover, the thorium-229 isotope features a nuclear transition 
proposed for a novel optical clock~\cite{Peik_2003,Peik_2021,Peik_2021_2}.  While the first resonant excitation of the clock transition has been achieved in a $^{229}$Th-doped CaF$_2$ crystal~\cite{Tiedau24}, followed by excitation in a doped LiSrAlF$_6$ crystal~\cite{Elwell24}, presently the search is on for the laser excitation with $^{229}$Th$^{3+}$ ions confined in a Paul trap, mitigating effects from the solid state environment for ultimate precision.

Molecular ions such as ThF$^+$ show an electric dipole moment and rotational degrees of freedom. 
For polar molecular ions, internal molecular states couple to the quantized modes of joint vibrations in the trap~\cite{HudsonPRA2018,Campell_PRL_2020}. Sufficiently strong dipole-phonon coupling make the internal (e.g., rotational) states of the trapped molecule amenable to quantum logic spectroscopy \cite{Campell_PRL_2020} or sympathetic cooling \cite{Leibscher26}. When the transition frequency between the internal molecular states is similar to the trap frequency, the coupling becomes resonant. This raises the question when such resonances occur. For example, rotational levels split by fine structure in CaO$^+$~\cite{Campell_PRL_2020,Qi_2024} were found to be resonant to trap frequencies of the order of a few MHz; similarly, purely rotational transitions in polyatomic asymmetric top molecular ions~\cite{Leibscher26}. In general, the answer depends on both, the masses and the molecular structure.

In this work, we investigate conditions for resonant dipole-phonon coupling for trapped molecular ions. To this end, we forego a truncation of the rotational state space~\cite{HudsonPRA2018,Campell_PRL_2020} to a few, pre-selected levels that couple strongly. Instead, we derive a rigorous model for the rotational motion of a co-trapped molecular ion that can be described as a rigid linear rotor. This applies in particular to diatomic molecular ions. Focusing, on ThF$^+$ as an example, we fully account 
also for the hyperfine interaction. This allows us to investigate resonant dipole-phonon coupling between the normal modes of the ion trap. Its hallmark is a splitting of the molecular levels, and we 
discuss how this can be detected using sideband-resolved laser (quantum logic) spectroscopy. 

The manuscript is organized as follows: We describe the theoretical framework in Sec.~\ref{sec:model}, starting with the model for dipole-phonon coupling in Sec.~\ref{subsec:model}.  We recap the calculation of vibrational common modes in  ion crystals with mixed species as well as the coupling of rotational degrees of freedom with the trap modes, before deriving the condition for resonant dipole-phonon coupling. In Sec.~\ref{subsec:diatomic} we describe the angular momentum states including hyperfine couplings and the resulting selection rules for dipole-phonon coupling for ThF$^+$. 
We present our results in Sec.~\ref{sec:results}, starting with the rotor-mass dependence on the frequency shifts and energy splittings in Sec.~\ref{subsec:generalfindings}. Conditions for resonant dipole-phonon coupling between hyperfine states of ThF$^+$ leading to an energy-splitting in the order of kHz for coupling to radial trap modes are discussed in Sect.~\ref{subsec:resonant_coupling_ThF}. Moreover, we discuss the experimental feasibility to detect this coupling with state-to-the-art methods in Sec.~\ref{subsec:spectroscopy} before concluding in Sec.~\ref{sec:outlook}.

\begin{figure}[tb]
	\includegraphics[width=1.0\linewidth]{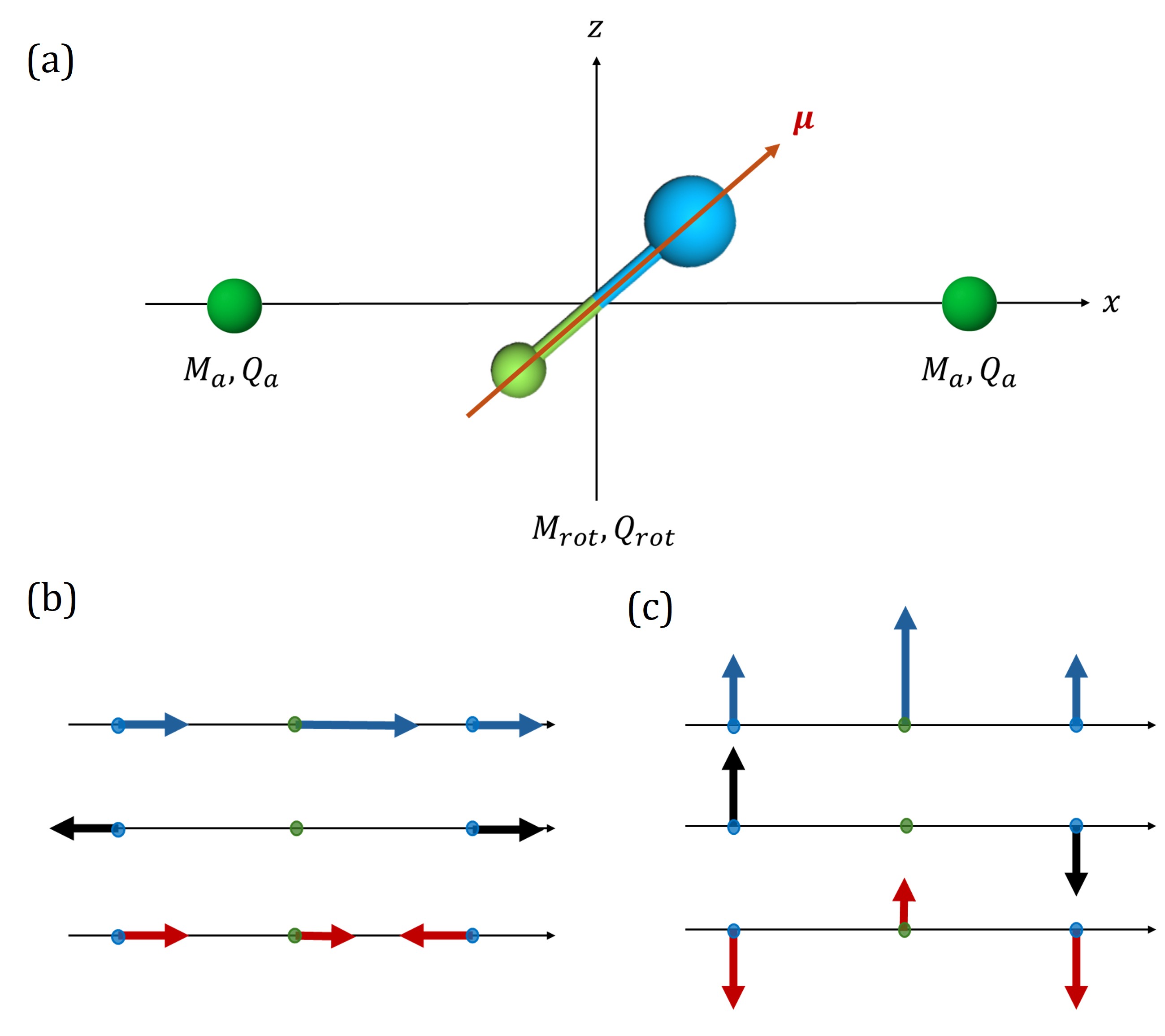} 
	\caption{Sketch of the setup: (a) We consider a linear crystal of trapped ions with a diatomic molecular ion at center. Atomic ions of charge $Q_a$ and mass $M_a$ are trapped in a harmonic Paul potential together with the charged rotor at the center. The rotor is described by its mass $M_{rot}$, charge $Q_{rot}$, and dipole moment $\mu$. 
     Different degrees of common vibrations couple to the rotational degree of freedom in the case of axial modes (b) or radial modes (c). The direction of eigenvectors is indicated by the colored arrows, namely for the center-of-mass (blue), breathing (black) and egyptian mode (red) in (b) and center-of-mass (blue), rocking (black) and zigzag mode (red) in (c).}
	\label{fig:schema}
\end{figure}

\section{Theoretical framework}
\label{sec:model}

\subsection{Model of a molecular ion coupled to the trap motion}\label{subsec:model}
We consider a linear Paul trap. The charged particles are trapped dynamically in $y$ and $z$ (radial) direction by an alternating radio-frequency, which generates a harmonic pseudo potential, and by a static harmonic electric potential in the $x$ direction. We assume the typical experimental situation, where the $x$-confinement is weaker as compared to the radial confinement. Moreover, we assume an anisotropic radial trap potential. This situation leads to a linear arrangement of trapped particles in the $x$-direction, coined as "linear crystal". In general, such crystals may contain many atomic and/or molecular ions, or charged clusters or nano-particles. This crystal with $N$ particles features $3N$ common modes of vibration, where the $N$ modes for each direction are uncoupled to any other.
In Fig.~\ref{fig:schema} (b) and (c) we sketch the modes in axial and in the radial direction. In the following we focus on the essential model with only three trapped objects: two atomic ions with mass $M_a$ and a molecular ion with mass $M_{rot}$ placed at the center of the crystal, as depicted in Fig.~\ref{fig:schema} (a). Note, that the loading and reordering of mixed ion crystals to achieve a specific arrangement has been experimentally demonstrated, e.g. in Ref.~\cite{zaw2024,kau2017,Spl09}.
The Hamiltonian describing the trap motion of the ions, i.e. the phonons, is given by
\begin{equation}
H_{ph} = \sum_p \omega_p \left( a_p^\dagger a_p + \frac{1}{2} \right),
\end{equation}
where $a_p$, $a_p^\dagger$ are the ladder operators of the normal modes. The harmonic approximation is well justified~\cite{Mar03}.
In the harmonic approximation, the modes in $y$ and $z$ direction are independent of each other, and for simplicity, we consider only the radial modes in $z$ direction. The frequency $\omega_p$ of any of the normal modes depends on the trap frequencies $\omega_x$ and $\omega_z$, and is obtained from the diagonalization of the Hessian matrix \cite{James_ApplPhys_1998,James_PRL_2000}. 
 Linear equilibrium configuration of the ions is assumed because the axial confinement is much weaker as compared to the radial one with $\omega_z \gg \omega_x$ \cite{James_PRL_2000}. 
The normalized eigenvectors of the Hessian matrix, ${\bf b}^{(p)} = (b_1^{(p)}, b_2^{(p)},b_3^{(p)})$, describe the collective motion of the trapped particles for each normal mode $p$, with $b_1^{(p)}$ and $b_3^{(p)}$ denoting the displacement of the atoms from equilibrium and $b_2^{(p)}$ the displacement of the center-of-mass of the molecule, cf. Fig.\ref{fig:schema} (b) and (c).
The displacement of the particles is depicted in Fig.~\ref{fig:schema} (b) for the axial normal modes and in
panel (c) for the radial modes. Note that for two modes, namely the axial breathing mode and the radial rocking mode (black arrows in 
Figs.~\ref{fig:schema} (b) and (c)), the motion of the atoms is decoupled from the rotor \rosi{which} is not displaced from the equilibrium, while for the other modes, the motion of all three particles is coupled. This very specific feature will be instrumental to selectively sense even small coupling strengths.

Inside the trapped atomic ion crystal, we now consider a trapped linear rotor with electric dipole moment. The rotor of mass $M_{rot}$ is charged with $Q_{rot}$ such that it is trapped in the Paul potential together with the atomic ions. The vibrational modes of mixed ion crystals featuring different $M/Q$-values has been worked out for two-ion crystals \cite{MORIGI2001,WUEBBENA2012}. 

Due to the rotational degrees of freedom, the Hamiltonian includes a term $H_{mol}$. 
The eigenstates of $ H_0 = H_{ph} + H_{mol}$ are the product states $\ket{\{n_p \},j} = \ket{n_1} \otimes ... \otimes \ket{n_6} \otimes \ket{j}$, where $n_p=0,1,2,...$ for $p=1,...,6$ are the quantum numbers of the normal modes and $j$ indicates 
the angular momentum quantum numbers of the molecular ion. The corresponding energy eigenvalues are
\begin{equation}
E_{\{n_p\},j} = \sum_p \hbar \omega_p \left( n_p + \frac{1}{2} \right) + E_j^{mol}
\label{eq:E0}
\end{equation}  
The molecules electric dipole moment ${\bm \mu}$ interacts with the electric field  ${\bf E}$, see Fig.~\ref{fig:schema}(a), at the center of mass position of the rotor ${\bf r}_{rot}$. The interaction energy reads as $H_{dp}=-{\bm \mu} \cdot {\bf E}({\bf r}_{rot})$. 
For each normal mode, the displacement of the rotor is given by ${\bf r}_{rot}=d_{rot} {\bf e}$ with ${\bf e} = {\bf e}_x$ for axial and ${\bf e} = {\bf e}_z$ for radial modes, where ${\bf e}$ is the direction of the electric field ${\bf E}$,
and~\cite{Campell_PRL_2020}
\begin{equation}
d_{rot} = \sqrt{\frac{\hbar}{2 M_{rot} \omega_p}} b_2^{(p)} \left ( a_p + a_p^\dagger \right).
\end{equation}
Determining the field strength by taking the gradient of the (harmonic) trap potential, the dipole interaction can be decomposed into components describing the interaction with a single normal mode, $H_{dp} = \sum_p H_{dp}^{(p)}$  with \cite{Campell_PRL_2020}  
\begin{eqnarray}
 H_{dp}^{(p)} =  {\cal E}_0^{(p)} \left( a_p + a_p^\dagger \right) {\bm \mu} \cdot {\bf e}.
 \label{eq:dipole_int}
\end{eqnarray}
Here,
\begin{equation}
{\cal E}_0^{(p)} =  b_2^{(p)} \sqrt{\frac{\hbar}{2e^2} \omega_p^3 M_{rot}}
\label{eq:field_strength}
\end{equation}
describes the scalar part of the electric field.  

In general, the eigenvalues of the Hamiltonian 
$H_0+ H_{dp}$ with $H_0=H_{ph}+H_{mol}$ can be obtained by numerical diagonalization. 
Since the dipole interaction is typically small compared to the eigenvalues of $H_0$, one can 
\rosi{approximate} the energy as
\begin{equation}
E^{dp}_{\{ n_p \},j} = E_{ \{n_p \},j} + \sum_p \Delta E_{n_p,j}^{(p)},
\label{eq:Edp}
\end{equation}
and calculate the modification $\Delta E_{n_p,j}^{(p)}$ 
 due to this interaction with normal mode $p$ in second order perturbation theory as
\begin{equation}
    \Delta E_{n_p,j}^{(p)} = \sum_{n_p',j'} \frac{ \left |\bra{n_p',j'} H_{dp}^{(p)} \ket{n_p,j}
    \right |^2}
    {E_{n_p,j} - E_{n_p',j'}}
    \label{eq_splitting_general}
\end{equation}
with $H_{dp}^{(p)}$ allowing transitions with $n_p' = n_p \pm 1$ and dipole allowed rotational transitions. 
Here, $\ket{n_p,j} = \ket{n_p} \otimes \ket{j}$ denotes the product state between a single normal mode $p$ and the rotational state. 
In this regime, despite  $n_p,j$ not being good quantum numbers, we can still use them to label the eigenvalues of $H_0+H_{dp}$.
If the denominator becomes zero, strong resonant interaction with a single normal mode $p$ is dominant and causes an energy splitting of the corresponding eigenvalues. In the non-resonant case, where the normal modes are sufficiently detuned from resonance condition, contributions from all normal modes lead to an overall shift of the eigenenergies. 

Close to resonance condition, contributions from other normal modes except the resonant mode $p$ can be neglected and the energy splitting due to dipole interaction becomes
\begin{equation}
\Delta E_{n_p,j}^{(p)} \approx \frac{ n_p |{\cal E}_0^{(p)}|^2}{E_{n_p, j}-E_{n_p-1,j'}}
  |\bra{j} {\bm \mu} \cdot {\bf e} \ket{j'}|^2.
\label{eq:splitting}
\end{equation}

The energy splitting is thus large if the energy gap between two molecular states, 
$E^{mol}_{j'}-E^{mol}_{j}$ is comparable to the normal mode frequency $\omega_p$, which is of the order of a few MHz. 
At exact resonance condition, a perturbative derivation is not possible and expression Eq.~(\ref{eq:splitting}) fails. In this case the couplings to all other states is negligible and  the energy splitting $\Delta E_{n_p,j}^{(p)}$ for $n_p > 0$
can be determined by diagonalizing
\rosi{the Hamiltonian of the}
 two level system with degenerate states $\ket{n_p,j}$, $\ket{n_p,j'}$, which results in 
\begin{eqnarray}
 \label{eq:splitting_resonant}
 \Delta E_{n_p,j}^{(p)} &=& \bra{n_p,j} H_{dp}^{(p)} \ket{n_p-1,j'} \nonumber \\
 &=& \sqrt{n_p} {\cal E}_0^{(p)} \bra{j}{\bm \mu} \cdot {\bf e} \ket{j'}.
\end{eqnarray}
We evaluate the rotational transition matrix element $\bra{j}{\bm \mu} \cdot {\bf e}  \ket{j'}$ for a diatomic molecule in Section \ref{subsec:diatomic}.
According to Eqs.~(\ref{eq:field_strength}), resonant energy splitting is proportional to $\omega_p^{3/2}$,
large normal mode frequencies thus lead to large energy splitting. The dependence of the energy splitting or shift on the rotor mass is less obvious since the coupling strength ${\cal E}_0^{(p)}$ depends on $M_{rot}$ directly and via $\omega_p$. Moreover, dipole interaction is also proportional to $b_2^{(p)}$, so that the interaction only occurs for normal modes for which $b_2^{(p)} \neq 0$, i.e. for which the center of mass of the rotor moves out of its equilibrium position.

Instead of directly observing the trapped rotor, its quantum states can be monitored by employing side-band spectroscopy on the atomic ions and apply and adapt some sort of quantum logic spectroscopy. Observation of the rotor thus relies on an efficient coupling of normal mode eigenvectors to the atomic ions and the trapped rotor. Measuring the dipole coupling via sideband spectroscopy thus also requires non-zero displacement of the atoms, i.e. $|b_1^{(p)}| = |b_3^{(p)}| \neq 0$.
Sideband spectroscopy of one of the atomic ions measures the frequency between the ground and excited state 
$\omega_l = \omega_a + \omega_p + \Delta \omega_p$, where $\hbar \omega_a$ is the energy difference between the ground and excited atomic state, $\omega_p$ is the normal mode frequency of the ionic crystal without dipole interaction and 
\begin{equation}
\Delta \omega_p = (\Delta E^{(p)}_{n_p=1,j} - \Delta E^{(p)}_{n_p=0,j}) / \hbar \approx \Delta E_{n_p=1,j}/\hbar.
\label{eq:frequency_shift}
\end{equation}
If the energy splitting $\Delta E^{(p)}_{n_p=1,j}$ results from resonant dipole-phonon coupling, we can neglect the contribution of 
$\Delta E^{(p)}_{n_p=0,j}$ since the ground motional and rotational state cannot couple resonantly and its energy shift is there for much smaller than the resonant part. 
With state of the art sideband spectroscopy, frequency shifts as small as several tens of Hz can be resolved.

\subsection{Angular momentum states and rotational transition matrix elements of a diatomic molecule: the hyperfine structure of ThF$^+$} \label{subsec:diatomic}
In the following, we discuss how the energy splitting due to the dipole-phonon interaction, Eq.(\ref{eq:splitting_resonant}) depends on the angular momentum of the trapped ion, assuming that the molecular ion is in its electronic and vibrational ground state. The total angular momentum includes molecular rotation, electronic orbital angular momentum and spin as well as the nuclear spins.
For an angular momentum state $\ket{j}=\ket{F,M_F}$, the dipole transition matrix elements are
  $\bra{F, M_F} {\bm \mu} \cdot {\bf e}_z \ket{F',M_F'} = \bra{F, M_F} T_0^1({\bm \mu}) \ket{F',M_F'}$
where $T_0^1({\bm \mu})$ is the spherical rank $1$ tensor of the dipole moment given in the space fixed frame. The Wigner-Eckart theorem states that \cite{Brown_Carrington}
\begin{eqnarray}
\label{eq:dipole_reduced_matrix_element} 
   &&\bra{F, M_F} T_0^1({\bm \mu}) \ket{F',M_F'} = \\
   && (-1)^{F-M_F}
    \begin{pmatrix}F&1&F'\\-M_F&0&M_F'\end{pmatrix} 
    \langle{F} \Vert T_0^1({\bm \mu}) \Vert F' \rangle,
    \nonumber
\end{eqnarray}
where $\langle{F} \Vert T_0^1({\bm \mu}) \Vert F' \rangle$ is the reduced matrix element. 
While the orientational selection rule $\Delta M_F=0$, which is implied in the Wigner-3j symbol, is independent of the specific composition of the total angular momentum ${\bf F}$, the evaluation of the reduced matrix element depends on the angular-momentum components contributing to ${\bf F}$.

For a rigid linear rotor with ${\bf F}={\bf R}$ and $R=0,1,2,..$
being the rotational quantum number, the reduced matrix element is given by
\begin{eqnarray}
  &&\langle{R} \Vert T_0^1({\bm \mu}) \Vert R' \rangle  = 
  \nonumber \\
  &&\mu (-1)^F \sqrt{(2R+1)(2R'+1)} \begin{pmatrix}R&1&R'\\0&0&0\end{pmatrix},
  \nonumber
\end{eqnarray}
where $\mu = |{\bm \mu}|$ is the
\rosi{modulus} of the dipole moment in the electronic and vibrational ground state of the molecule and only transitions with $\Delta R = \pm 1$ are allowed. 
The energy difference between rotational states that interact via dipole-phonon coupling is $E_R^{rot}-E_{R-1}^{rot}=2BR$. Since the rotational constant $B$ of a diatomic molecule is typically in the GHz regime, rotational transitions are far off resonant to the trap modes and they couple only non-resonantly. For heavy diatomic molecular ions like ThF$^+$, the energy shift due to the non-resonant coupling can be estimated to be of at most a few Hz.  
However, fine and hyperfine interactions 
split the rotational levels and enable dipole allowed transitions between states with energy differences on the order of MHz and thus facilitate resonant coupling to the trap modes. 

For heavy diatomic molecules the coupling of the angular momenta is often well described by Hund's case (c), see Ref.~\cite{Brown_Carrington}. In this case, the total electronic orbital and spin angular momenta ${\bf L}$ and ${\bf S}$ strongly couple and form the total electronic angular momentum ${\bf J}_e={\bf L}+{\bf S}$ with projection 
${\bm \Omega}$ along the internuclear axis. The total angular momentum (without nuclear spins) is then ${\bf J} = {\bf R} + {\bf J}_e$ with rotational energies $E_J^{rot}=B J(J+1)$. In this case, the reduced transition matrix element is  $\langle{F} \Vert T_0^1({\bm \mu}) \Vert F' \rangle =
\langle{J,\Omega} \Vert T_0^1({\bm \mu}) \Vert J', \Omega \rangle$ with
\begin{eqnarray}
   && \langle{J,\Omega} \Vert T_0^1({\bm \mu}) \Vert J', \Omega \rangle = \nonumber \\
  &&  \mu (-1)^{J-\Omega}\sqrt{(2J+1)(2J'+1) } \begin{pmatrix}J&1&J'\\-\Omega&0&\Omega\end{pmatrix}. \nonumber
\end{eqnarray}
For $\Omega=0$, only transitions with $\Delta J=\pm 1$ are allowed, which do not enable resonant coupling.
\changed{ For $\Omega \neq 0$, the molecular eigenstates occur in form of parity doublets
\begin{equation}
  \ket{J\Omega M p} = \frac{1}{\sqrt{2}} \left( \ket{J\Omega M} +p (-1)^J \ket{J-\Omega M} \right),  
  \nonumber
\end{equation}
with $p=\pm$
and dipole coupling occurs between states with opposite parity and is allowed for states with $\Delta J= \pm 1$ and $\Delta J=0$.} If the  rotational states are further split e.g. by hyperfine interactions, the latter can lead to resonant dipole-phonon coupling.

The details of the hyperfine interaction depend on the internal structure of the diatomic molecule. 
We consider in particular ThF$^+$, containing $^{229}$Th with $I_1=5/2$ and $^{19}$F with $I_2=1/2$, in its electronic ground state $^3\Delta_1$ \cite{Ng2022}, which has the projection quantum number $\Omega=1$ and thus $\Delta J=0$ transitions are electric-dipole allowed.  
The addition of the nuclear spins is considered independently, such that ${\bf F}_1 = {\bf J} + {\bf I}_1$ and ${\bf F} = {\bf F}_1 + {\bf I}_2$.
The basis of the hyperfine states is $\ket{J \Omega I_1 F_1 I_2 F M_F}$, and
the reduced matrix element \changed{in the basis of the parity eigenstates}
\rosi{gives} ~\cite{Brown_Carrington}
\begin{eqnarray}
  \label{eq:reduced_matrix_ThF}
   &&\langle J \Omega I_1 F_1 I_2 F \mp \Vert T^1({\bm \mu}) 
    \Vert J' \Omega I_1 F_1' I_2 F' \pm \rangle  = \nonumber \\
   && \mu (-1)^{F_1+F_1'+F'+I_1+I_2-\Omega}
   \sqrt{(2J+1)(2J'+1)} \nonumber \\
   && \times \sqrt{(2F_1+1)(2F_1'+1)(2F+1)(2F'+1)} \begin{pmatrix}J&1&J'\\-\Omega&0&\Omega\end{pmatrix}
   \nonumber \\
   && \times \begin{Bmatrix}J'&F_1'&I_1\\ F_1&J&1 \end{Bmatrix} \begin{Bmatrix}F_1'&F'&I_2\\ F&F_1&1 \end{Bmatrix}.
\end{eqnarray}
This determines the selection rules $\Delta J = 0,\pm 1$ for $\Omega=1$ as well as $\Delta F_1 = 0, \pm 1$
and $\Delta F = 0, \pm 1$.
For resonant dipole-phonon interactions, the relevant transitions are $\Delta J=0$, $\Delta F_1 =0$ and $\Delta F = \pm 1$.

\rosi{For ThF$^+$, the effective Hamiltonian describing the hyperfine interaction~\cite{Brown_Carrington} reads } 
\begin{equation}
H_{int}=H_Q+H_{IL}+H_{IS}+H_{IJ}+H_{II}.
\label{eq:hf_hamil}
\end{equation}
The nuclear electric quadrupole interaction $H_Q$ as well as the interaction between the nuclear spins of Th and F with the electronic angular momenta, $H_{IL}$ and $H_{IS}$, 
are the dominant terms. The nuclear spin-molecular rotation interaction $H_{IJ}$, 
\rosi{and the interaction between the two nuclear spins}
$H_{II}$, are several orders of magnitude  weaker. 

We describe the explicit effective hyperfine Hamiltonian $H_{int}$ and the computation of corresponding matrix elements in Appendix \ref{appendix:hyperfine}.  
The hyperfine-splitting is obtained by
 \rosi{solving the Schrodinger equation associated to} the Hamiltoninan $H_{mol} = H_{rot} + H_{int}$ 
\rosi{performing a basis set expansion in terms of}
the $\ket{J\Omega I_1 F_1 I_2 F M_F}$ basis. 
The resulting hyperfine spectrum of ThF$^+$ is discussed in Section \ref{sec:results}. 
In general, the basis states $\ket{J\Omega I_1 F_1 I_2 F M_F}$ are not eigenstates of $H_{mol}$. However, for ThF$^+$, the mixing between different basis states is so small, that we can approximate the eigenstates by the basis states and directly use Eq.~\ref{eq:reduced_matrix_ThF} to calculate the dipole transition matrix elements. 

\section{Results}
\label{sec:results}

We start by identifying, for singly charged ions, which normal modes couple to the rotation. Then we focus on resonant dipole-phonon coupling for hyperfine states of 
ThF$^+$ ions and finally, we discuss how energy shifts due to the coupling can be observed by high resolution sideband spectroscopy. 
  
\subsection{Coupling between rotation and normal modes of the three-ion crystal}
\label{subsec:generalfindings}
\begin{figure}
    \includegraphics[width=\linewidth] {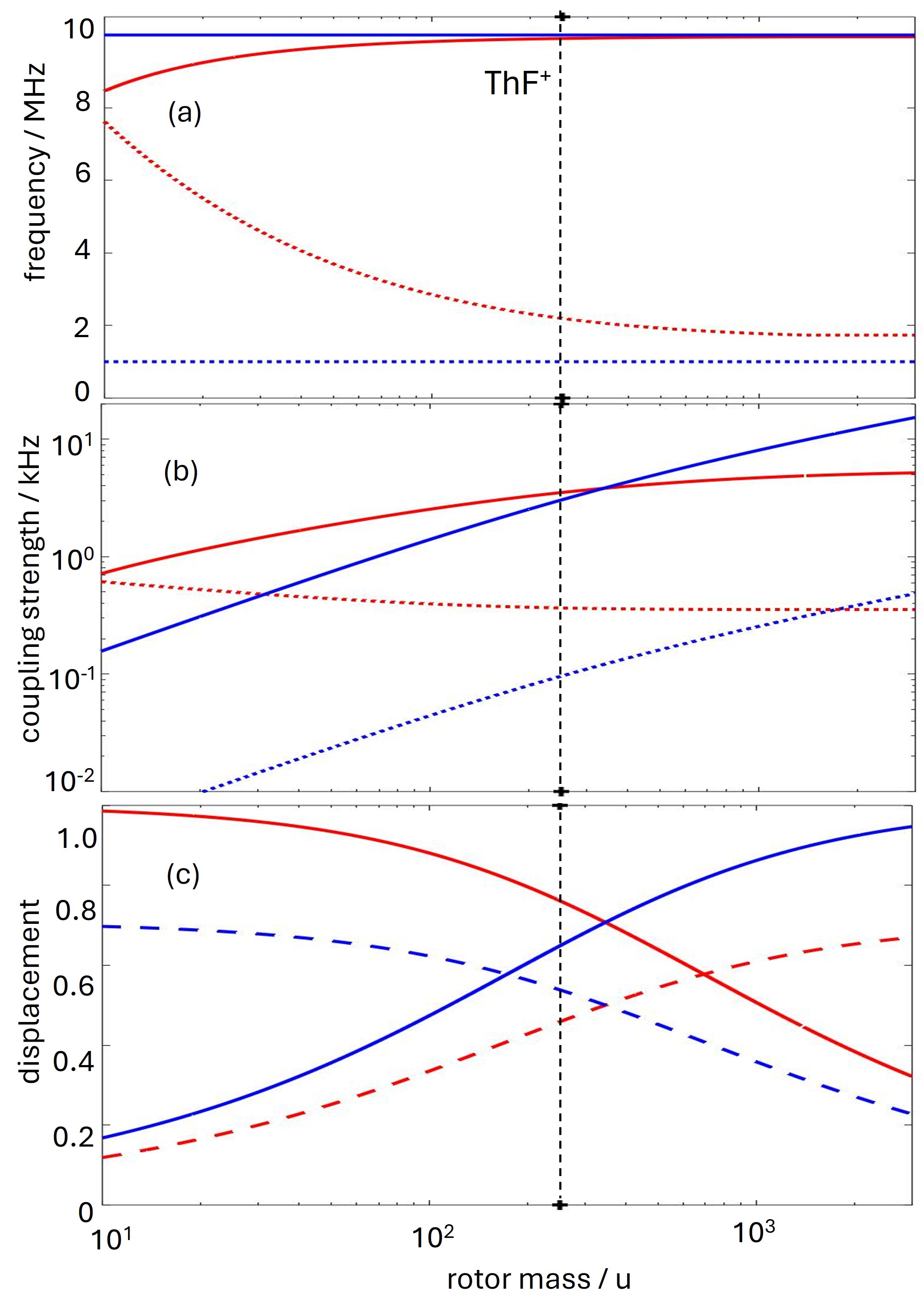}
	\caption{
    (a) Normal mode frequencies for for the radial center-of-mass mode (solid blue) and zigzag mode (solid red) as well as for the axial center-of-mass (dotted blue) and Egyptian mode (dotted red) modes.
    (b) Dipole-phonon coupling strength 
    $ |{\cal E}_0^{(p)} \mu / h| $ for a dipole moment of $\mu=1$ D for the same modes, and (c) displacement $|b_1^{(p)}|=|b_3^{(p)}|$ of the trapped  atomic ions (dashed lines) and displacement $|b_2^{(p)}|$ of the center-of-mass of the molecular ion
    \rosi{(solid lines)}. The radial and axial center-of mass modes (red curves) as well as the Egyptian and zig-zag modes (blue curves) have the same displacement. 
    The vertical dashed line indicates $M_{rot}=248$ u, i.e. the mass of ThF$^+$. We consider singly charged atomic ions of mass of $M_a=173$ u, 
     the trap frequencies are $\omega_z=1$ MHz and $\omega_y=10$ MHz. 
     }
	\label{fig:normal_modes_motion}
\end{figure}

In the following, we analyze how the axial and radial center-of-mass modes and the egyptian and zig-zag modes, see~Figs.~\ref{fig:schema}(b) and (c),  couple to the dipole of the molecular ion as a function of the mass of the molecule.
To obtain the normal mode frequencies $\omega_p$ and the corresponding (normalized) eigenvectors ${\bf b}^{(p)}$, we numerically diagonalize the corresponding Hessian matrix \cite{James_ApplPhys_1998}. 
Figure~\ref{fig:normal_modes_motion} (a) shows the normal mode frequencies as function of the rotor mass. 
Except for very light molecules, the frequencies of the radial modes (solid lines) are much larger than the frequencies of the axial modes (dotted lines). Since the coupling strength
${\cal E}_0^{(p)}$ scales with $\omega_p^{3/2}$, see Eq.~(\ref{eq:field_strength}), the radial modes couple stronger to the molecular dipole than the axial modes, as it can be seen Fig.~\ref{fig:normal_modes_motion} (b). Moreover, the coupling strength increases with the rotor mass. Only coupling to the egyptian mode (dotted red line) is strongest for light molecules since here, the normal mode frequency decreases rapidly with the rotor mass. For ThF$^+$, the frequencies of the zig-zag and radial center-of-mass modes are almost the same; coupling to the zig-zag mode is slightly stronger since the displacement of the rotor, $|b_2^{(p)}|$, shown in panel (c) 
of Figure~\ref{fig:normal_modes_motion}, is larger for the zig-zag mode than for the center-of-mass mode. 

Using quantum logic spectroscopy to measure or manipulate the internal states of the molecular ion via dipole-phonon coupling requires that also the atoms are coupled strongly to the trap modes. The displacement of the atoms from their equilibrium position indicates their coupling to the trap modes. This is shown in Fig.~\ref{fig:normal_modes_motion} (c) with dashed lines for the components $|b_1^{(p)}| = |b_3^{(p)}|$ of the eigenvector $b^{(p)}$. Note that the axial and radial center-of-mass modes have the same eigenvectors (blue lines), and so do the egyptian and zig-zag modes (red lines). For ThF$^+$, both atomic and molecular ions couple strongly to the trap modes. For molecular ions that are much lighter or heavier than the atomic ions (here Yb$^+$ with $M_a=172$ u), the motion of atoms and molecules decouple, but it should be noted that for a large range of rotor masses, the motion of all trapped particles is coupled.

In summary, coupling between the (radial) motion of the trapped atomic ions and the rotation of the molecules with a coupling strength in the order of kHz can be achieved for a large range of polar molecular ions. The results shown 
\rosi{here} hold for any (singly charged) polar molecule with masses between 10 to a few thousand u. However, for resonant dipole-phonon coupling, also the resonance condition between rotational transitions and the normal mode frequencies has to be fulfilled, and this depends strongly on structure of the molecular ion. In Ref.~\cite{Leibscher26} it has been shown that the resonance condition can be fulfilled for a large range of asymmetric top molecules. In the following, we discuss resonant dipole-phonon coupling for diatomic molecular ions, in particular for ThF$^+$, where coupling between rotation, electronic angular momenta and nuclear spins is essential to have resonant dipole-phonon coupling. 

\subsection{Resonant dipole-phonon coupling for diatomic molecular ions}
\label{subsec:resonant_coupling_ThF}
\begin{figure}
   \includegraphics[width=1.0\linewidth]{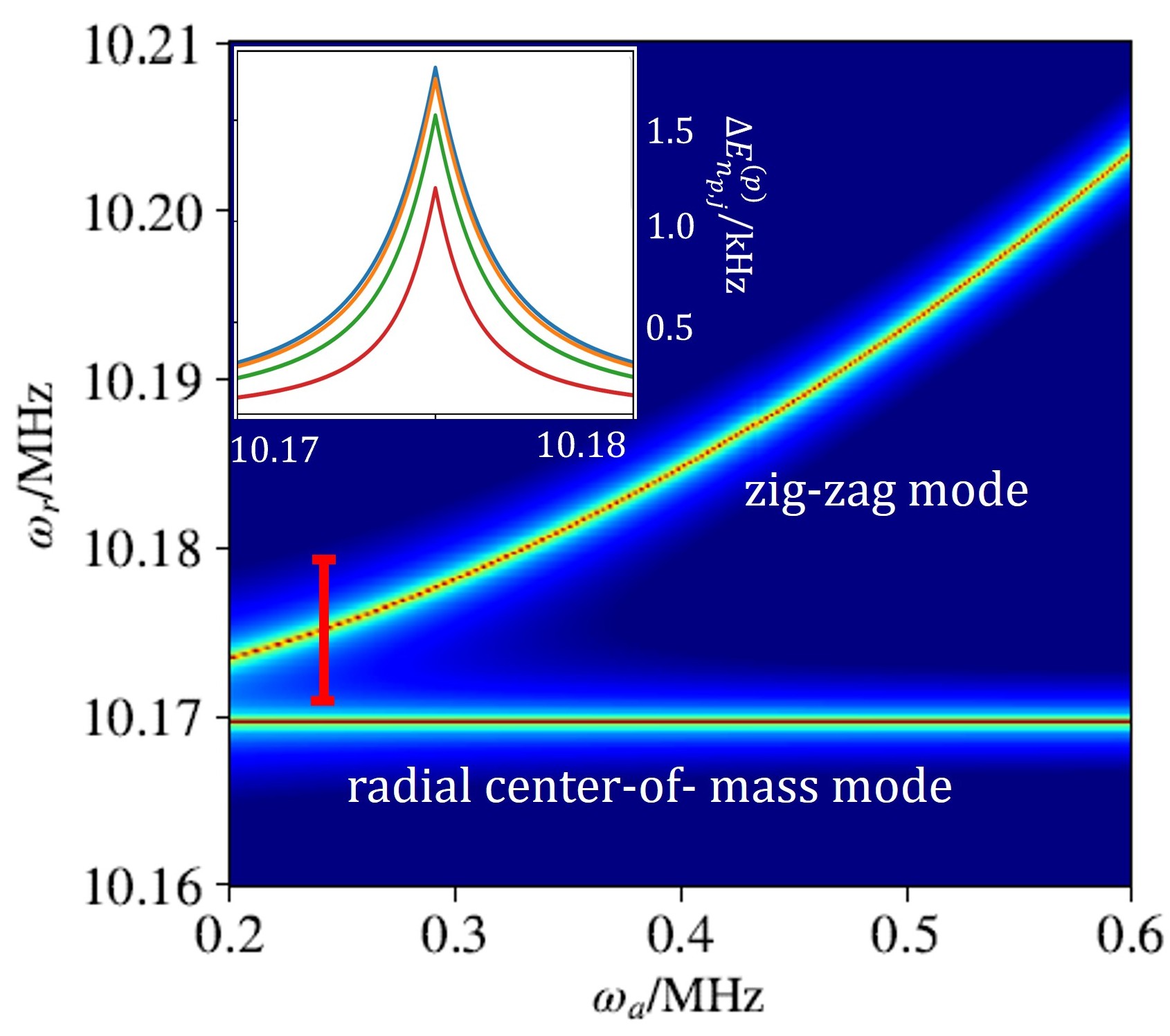}
	\caption{Resonant dipole phonon coupling in a Coulomb crystal with two Yb$^+$ and a ThF$^+$ ion.  Energy splitting $\Delta E^{(p)}_{n_p,j}$ as a function of the radial and axial trap modes for the zig-zag-mode and radial center-of-mass mode for $n_p=1$ and $M_F=0$. The inset shows  $\Delta E^{(p)}_{n_p,j}$ as a function of the radial trap frequency for $\omega_a=0.24$ MHz for the zig-zag mode, as indicated by the red bar in the 2D plot. Here the energy splitting is shown for $M_F=0$ (blue) and $|M_F|=1,2,3$ (orange, green, red).}
	\label{fig:resonant_coupling}
\end{figure}

The splitting of the rotational levels of ThF$^+$ due to the hyperfine interactions Eq.(\ref{eq:hf_hamil}) for $J=1$ is shown in Figure \ref{fig:spectrum_ThF}. The relevant terms together with the molecular parameters are described in detail in Appendix \ref{appendix:hyperfine}.  
\begin{figure}
   \includegraphics[width=1\linewidth]{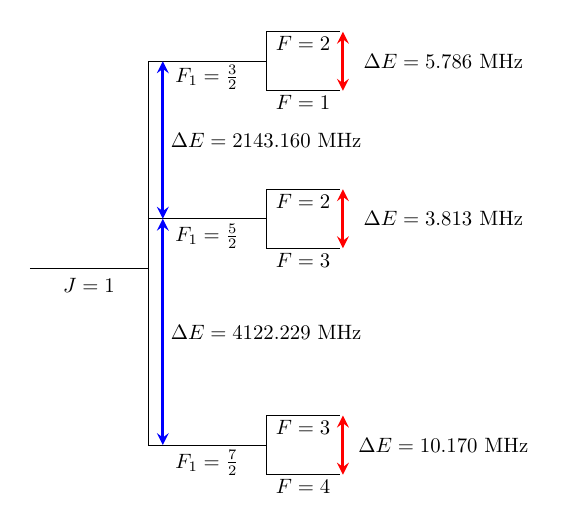}
	\caption{Hyperfine states of $^{229}$Th$^{19}$F$^+$ in the electronic ground state for $J=1$. The hyperfine interaction Hamiltonian and molecular parameters are discussed in Appendix \ref{appendix:hyperfine}, where we also show the spectrum for $J=2$.}
	\label{fig:spectrum_ThF}
\end{figure}
The splitting of $J=1$ into three $F_1$-levels results mainly from the interaction of the Th nuclear spin with the electronic angular momenta and from the quadrupole interaction, while 
splitting of the $F_1$ states into pairs of $F$ states comes from the interaction of the fluorine nuclear spin with the electronic angular momenta.
The remaining terms of Eq.~(\ref{eq:hf_hamil}) slightly modify the energy splittings, see Appendix~\ref{appendix:hyperfine} for details. 
While the $\Delta F_1=1$ transitions are far off resonance to any of the normal modes, transitions with $\Delta F_1=0$ and $\Delta F=1$ are in the order of a few MHz.  One of the motional frequencies can thus be tuned such that one of the hyperfine transitions is resonant to a trap mode. Since large normal mode frequencies $\omega_p$ cause strong dipole-phonon coupling, we expect that the dipole-phonon coupling is particular strong if one of the radial normal modes couples resonantly with the $F$ states with the largest gap ($10.17$ MHz), namely between the states $J=1$, $F_1=7/2$, $F=3$ and $4$, cf. Fig.~\ref{fig:spectrum_ThF}. The energy splitting due to resonant dipole-phonon coupling, $\Delta E_{n_p,j}^{(p)}$, is shown in Fig.~\ref{fig:resonant_coupling} as a function of the axial and radial trap frequencies. 
We identify a strong enhancement of the dipole-phonon energy splitting for trap frequencies, for which the resonance condition
$\omega_p = E^{rot}_{j'}-E^{rot}_{j}$ is fulfilled. 
The resonance profile is shown as inset of Fig.~\ref{fig:resonant_coupling} as a function of the radial trap frequency across the \rosi{marked} range (red bar) in the 2D plot. The maximal splitting at resonance condition is approximately $1.5$ kHz with a slight dependence of the dipole-phonon coupling strength on the orientational quantum number $M_F$. The dipole coupling is maximal for $M_F=0$ and smallest for $M_F=F$, jet all pairs of resonantly coupled hyperfine states display an energy-splitting of about $1$ kHz. 

The trap can also be tuned such that a normal mode is resonant to one of the other $\Delta F=1$ transitions. The maxima of the resulting energy splittings (for $M_F=0$) are reported in Table \ref{table:energy_splittings} for interaction with the radial center-of-mass mode and with the zig-zag mode.

\begin{table*}
\centering
\begin{tabular}{|c|c|c|c|c|c|}
\hline
Transition ($M_F=0$) & $\omega_p$ / MHz &  $\Delta E_{n_p,j}$ / Hz & $\Delta E_{n_p,j}$ /Hz  & 
$\Delta E_{n_p,j}$ /Hz  & $\Delta E_{n_p,j}$ /Hz\\
 & & zig-zag & radial c.o.m. & egyptian & axial c.o.m. \\ \hline
 $|1,\frac{7}{2},4\rangle\longleftrightarrow|1,\frac{7}{2},3\rangle$ & 10.170 & 1760 & 1490 & - & -\\\hline
$|1,\frac{5}{2},3\rangle\longleftrightarrow|1,\frac{5}{2},2\rangle$ &  3.813 & 162 & 137 & 81 & - \\\hline
 $|1,\frac{3}{2},1\rangle\longleftrightarrow|1,\frac{3}{2},2\rangle$ & 5.786 & 1058 & 895 & - &-\\\hline
 $|2,\frac{9}{2},5 \rangle\longleftrightarrow|1,\frac{9}{2},4\rangle$ & 6.552 & 472 & 400 & - & -\\\hline
 $|2,\frac{7}{2},4\rangle\longleftrightarrow|2,\frac{7}{2},3\rangle$ & 4.167 & 222 & 188 & 111 &-\\\hline
 $|2,\frac{5}{2},3\rangle\longleftrightarrow|2,\frac{5}{2},2\rangle$ & 2.349 & 78 & 66 & 39 & 33\\\hline
 $|2,\frac{3}{2},2\rangle\longleftrightarrow|2,\frac{3}{2},1\rangle$ & 0.560 & 3.5 & 3.0&1.8 & 1.5\\\hline
 $|2,\frac{1}{2},0\rangle\longleftrightarrow|2,\frac{1}{2},1\rangle$ &  3.269 & 499 & 422&250 &-\\\hline
\end{tabular}
\caption{Maximal energy splitting due to resonant dipole-phonon coupling for $M_F=0$ and $n_p=1$ between the hyperfine states \changed{$\ket{j}=\ket{J \Omega I_1 F_1 I_2 F M_F\pm}$ and
$\ket{j'}=\ket{J \Omega I_1 F_1 I_2 (F-1) M_F \mp}$. For simplicity we only denote
$\ket{j}=\ket{J,F_1,F}$ and $\ket{j'}=\ket{J,F_1,F-1}$ in the table, since all other quantum numbers are fixed.} The splitting is given for the zig-zag mode, radial center-of-mass mode, egyptian mode and axial center of mass mode.}
\label{table:energy_splittings}
\end{table*}

The table shows that for all transitions, resonant coupling to the zig-zag mode is always stronger than for the center-of-mass mode, which is in accordance to Fig.~\ref{fig:normal_modes_motion} (b).
It can also be seen that for all hyperfine transitions that are of the order of a few MHz, dipole-phonon splittings of a few hundred Hz up to a kHz can be obtained. 
Since a linear trap is much less stiff in the axial direction than in the radial one, the axial normal mode frequencies are much smaller than the radial ones, cf. Fig.~\ref{fig:normal_modes_motion} (a). As a result, the dipole-phonon coupling is much weaker. Moreover, not all pairs of hyperfine $F$-states can be coupled resonantly to the axial modes. 
The last two columns of Table \ref{table:energy_splittings} show the energy splittings for those transitions, that can be coupled resonantly to the axial modes. Here, we find energy splittings of the order of a few tens of Hz. The egyptian mode has - for the same trap - a slightly larger frequency and coupling strength, as it can be seen in Fig.~\ref{fig:normal_modes_motion} (a) and (b). Coupling to this mode is therefore slightly larger, and resonance condition can be fulfilled for more transitions than for the axial center-of-mass mode. 

\subsection{Detecting dipole-phonon coupling with sideband spectroscopy}\label{subsec:spectroscopy}
The resonant energy splitting can be observed with sideband spectroscopy addressing the atomic ions. This requires a spectral resolution in the range of hundreds of Hz to kHz
for resonant coupling to the radial modes, and of several tens of Hz for coupling to the axial modes.
In terms of a practical protocol, and assuming that an experimental setup allowing for a resolution in the range of Hz is available, one would search for a mismatch between the {\em ratios} of the normal mode frequencies~\cite{James_ApplPhys_1998} e.g. for the axial center-of-mass, breathing, and Egyptian mode. Without dipole-phonon coupling, this frequency ratio is fixed in a harmonic trap potential and with linearized Coulomb interaction between the ions. This ratio can be determined by first accurately characterizing the trapping potential with a single atomic ion, and then deriving the desired normal mode frequencies by inserting the respective atomic and molecular masses. 
As all particles are confined at the RF null of a linear quadrupole field, no deviations from this frequency ratio are expected.  While a deviation from the expected frequencies has been observed for radial modes in a \textit{planar} ion crystal~\cite{KaufmannPRL2012}, in agreement with a accurate calculation using the Floquet-Lyapunov approach, for a \textit{linear} ion crystal and a harmonic axial potential, any ever-so-slight deviation from the predicted ratio would be indicating a rotor-phonon coupling.

If the spectral resolution is not sufficient to directly measure the frequency shift, one might employ heating rate measurements on the different modes. These techniques are established to characterize the properties of ion crystals that are parasitically coupled in their vibrational degrees of freedom to the trap surfaces~\cite{GLI2025,An2019,Hou2025}. Modes that couple to the rotor will show an increased motional decoherence rate, as in a mimicked measurement~\cite{Fol2023}, similar to the effect internal phonons~\cite{Hen24}. Such a measurement scheme would realize quantum logic 
sideband spectroscopy where heating 
is detected by a weakly coupled spectroscopy ion
~\cite{Gug15}, but here with the mode-specific coupling of the atomic ions and to the molecular ion rotor degrees of freedom. The goal would be to 
find a rotor-specific heating rate that can be selectively switched on and off by very small changes in the resonance condition for this coupling, enabled by changes of the trap control voltages. Common noise rejection techniques \cite{Chwalla2007} or designer ion pairs \cite{Roos2006,Kotler2014} may be employed to further push the detection limits for the rotor coupling.

\section{Summary and Outlook}\label{sec:outlook}

We have studied the dipole-phonon coupling for an exemplary three-ion chain consisting of a polar rotor in between two atomic ions, extending earlier work on such coupling~\cite{HudsonPRA2018,Campell_PRL_2020} to account for the full rotational state space.
Investigating how the normal-mode frequencies and dipole–phonon coupling strength depend on the mass of the molecular ion, we have found that polar molecular ions with masses ranging from a few tens to a few thousand u can achieve coupling strengths to radial modes on the order of kHz, with corresponding normal-mode frequencies in the MHz range.
Focusing on diatomic ions, we have included spin-orbit and hyperfine interactions in the description of the molecular ion. 
When these interactions are sufficiently strong, they split a given rotational level into a manifold of sublevels with  energy gaps of the order of a few MHz. Since these transitions are dipole-allowed, they lead to considerable dipolar coupling between rotational states and the trap normal modes. Our work provides the detailed derivation of the coupled spin-rotational structure for $^{229}$Th$^{19}$F$^+$, including the \textit{ab initio} calculation of the hyperfine parameters.
We have calculated the dipole-phonon coupling for ThF$^+$ co-trapped with two Yb$^+$ ions, identifying various feasible trap configurations for which the coupling becomes resonant. In this case, the dipole-phonon interaction results in energy level splittings in the order of kHz for coupling with radial trap modes and tens of Hz for coupling with axial normal modes. 
Such shifts should be observable with high-resolution sideband spectroscopy on the atomic ions, or decoherence measurements for the vibrational modes of the trap.

While this work considered singly charged diatomic ions, molecules in higher charge states can also be prepared and trapped~\cite{zulch:2022,Stricker_2025}. 
The higher charge increases the normal mode frequency, and multiply charged ions tend to possess larger dipole moments. As a result, significantly larger dipole-phonon coupling strengths may be expected for multiply charged ions, making them an interesting platform for future work. 
In addition to realizing quantum logic~\cite{HudsonPRA2018,Campell_PRL_2020}, resonant coupling of a rotor to the normal modes of an ion Coulomb crystal could also be employed for cooling the rotational degrees of freedom. A recently suggested protocol for large trapped molecular ions~\cite{Leibscher26} readily extends to the diatomic case with strong spin-orbit coupling, as present in $^{229}$Th$^{19}$F$^+$. It will enable more accurate quantum control of molecular ion states, implying exciting prospects for quantum information and optical clocks.

\begin{acknowledgments}
 We gratefully acknowledge financial support from the Deutsche Forschungsgemeinschaft through CRC 1319 ELCH and through the joint ANR-DFG CoRoMo project 505622963 / KO 2301/15-1, ANR-22-CE92-0077-01 and DFG Pr. TACTICa Nr. 495729045, and of the Spanish project PID2023-147039NB-I00. J.M.G.G. acknowledges financial support from the grant PRE2021-099603 funded by MICIU/AEI/10.13039/501100011033 and by ``ESF+". K.G. thanks the Fonds der Chemischen Industrie (FCI) for generous funding through a Liebig fellowship. FSK thanks R. Folman for discussions.
\end{acknowledgments}

\begin{appendix}
  \section{Hyperfine structure of ThF$^+$}
   \label{appendix:hyperfine}
   
   \begin{table}
  	\begin{center}
		\begin{tabular}{c|c|c|c|c|c|c }  
          \hline
		 	 $B$ & $eQq_{zz}$ &$A_\Vert (\mbox{Th})$ & $A_\Vert (\mbox{F})$ & $c(\mbox{Th})$ & $c(\mbox{F})$ & $\mu$ \\
             
			\hline
            
		    7347 & $-3687$ & $-2054$ & $-16$ & $-0.362$ & $-0.022$  & 3.4 \\
          \hline  
		\end{tabular}
	\end{center}
   \caption{Parameters for $^{229}$Th$^{19}$F$^+$ in the electronic ground state $^3\Delta_1$ in MHz computed at the level of 2c-ZORA-cGKS-BHandH. 
   } 
    \label{tab:parameter_ThF}
\end{table}

  In this appendix, we discuss the Hamiltonian $H_{mol}=H_{rot}+H_{int}$ of $^{229}$Th$^{19}$F$^+$ in its vibrational and electronic ground state, employing the basis states $\ket{J \Omega I_1 F_1 I_2 F M_F}$, i.e. Hund's case (c).

 We computed all relevant molecular parameters with a modified version
\cite{gaul:2020, zulch:2022, bruck:2023, colombojofre:2022} of a two-component program \cite{wullen:2010} based
on Turbomole \cite{ahlrichs:1989} within a quasi-relativistic complex generalized Kohn-Sham (cGKS)
framework including relativistic effects on the two-component zeroth order regular approximation
(2c-ZORA) level. We employed the hybrid
local density approximation (LDA) using the X$\alpha$ exchange
functional \cite{dirac:1930,slater:1951} and the VWN-5 correlation
functional \cite{vosko:1980} with 50\,\% Fock exchange by Becke (BHandH)
\cite{becke:1993}. Results of calculations with this functional have previously been found to be in reasonable agreement with more sophisticated methods such as
Coupled Cluster for molecular properties of heavy element compounds
\cite{gaul:2024,gaul:2024a}.
2c-ZORA was employed with a damped model potential to alleviate the gauge dependence
\cite{wullen:1998,liu:2002}. We employed atom-centered
Gaussian basis functions using the core-valence basis set of triple-$\zeta$
quality by Dyall (dyall.cv4z) \cite{dyall:2002, dyall:2006}, which was augmented with an additional set of 12 s and 6 p functions for the description of the wave function within the nucleus. These additional sets of
steep functions were composed as an even-tempered series starting at
$10^{9}\,a_{0}^{-2}$ and progressing by division by $1.5$ for s
functions and $2.5$ for p functions. 

The nuclear charge density distribution was
modeled as a normalized spherical Gaussian 
$\varrho_K \left( \bm{r} \right) = \frac{\zeta_K^{3/2}}{\pi ^{3/2}}
\mathrm{e}^{-\zeta_K \left| \bm{r} - \bm{r}_K \right| ^2}$ with $\zeta_K =
\frac{3}{2 r^2 _{\mathrm{nuc},K}}$. The root-mean-square radius
$r_{\mathrm{nuc},K}$ was chosen as suggested by Visscher and Dyall employing the 
isotopes $^{19}$F and $^{232}$Th \cite{visscher:1997}. The molecular structure and electronic densities were previously optimized at this level of theory in Ref.~\cite{gaul:2024a}. The equilibrium bond length was found to be $3.742\,a_0$, which is in reasonable agreement with the experimental value of $3.7571(5)\,a0$ \cite{Ng2022}. The error of the resulting rotational constant $7347\,\mathrm{MHz}$ is $<1\,\%$ compared to the experimental value of Ref.\,\cite{Ng2022} ($7288(2)\,\mathrm{MHz}$) and completely sufficient for the present purpose. We used this bond length and the electronic densities of Ref.~\cite{gaul:2024a} to compute the hyperfine structure parameters employing following expressions:
\begin{align}
e Q q_{zz} &= \frac{e^2Q}{4\pi\epsilon_0\,h}\left[-\Braket{\frac{2\frac{\partial^2}{\partial z_{\mathrm{Th}}^2}-\frac{\partial^2}{\partial x_{\mathrm{Th}}^2}-\frac{\partial^2}{\partial y_{\mathrm{Th}}^2}}{3}\sum\limits_i\phi_{\mathrm{Th}}(\bm{r}_i)}\right.\nonumber\\
&\left.+\frac{2 Z_{\mathrm{F}}}{\left|\bm{R}_{\mathrm{Th}}-\bm{R}_{\mathrm{F}}\right|^3}\right] \\
A_\parallel(X) &= \frac{e\,c\,\hbar\,\mu_0\,\gamma_X}{4\pi\,h\,\Omega} 
\Braket{\sum\limits_i
\frac{(\bm{r}_i-\bm{R}_X)\times\bm{\alpha}_i}{\left|\bm{r}_i-\bm{R}_X\right|^3}} \\
c(X) &= \frac{e\,c\,\hbar\,\mu_0\,\gamma_X\,B_e}{4\pi\,h} \Braket{\Braket{\sum\limits_i
\frac{(\bm{r}_i-\bm{R}_X)\times\bm{\alpha}_i}{\left|\bm{r}_i-\bm{R}_X\right|^3};\hat{\bm{J}}_\mathrm{e}}}\,,
\label{eq:nsmr}
\end{align}
where sums run over all electrons $i$ and the scalar potential from a spherical Gaussian charge-density distribution of the Th nucleus is $\phi_\mathrm{Th}(\bm{r})=\frac{\mathrm{erf}\left(\sqrt{\zeta_\mathrm{Th}}\left|\bm{r}-\bm{R}_\mathrm{Th}\right|\right)}{\left|\bm{r}-\bm{R}_\mathrm{Th}\right|}$. $\Braket{\hat{A}}$ denotes the electronic expectation value of operator $\hat{A}$ within the Born--Oppenheimer approximation, and $\Braket{\Braket{\hat{A},\hat{B}}}$ is the electronic linear response function associated with perturbations $\hat{A}$ and $\hat{B}$.
Here we used the magnetogyric ratios $\gamma_{^{229}\mathrm{Th}}=0.184(16)\,\mu_\mathrm{N}$, $\gamma_{^{19}\mathrm{F}}=5.257736(16)$ and the nuclear electric quadrupole moment $Q_{^{229}\mathrm{Th}}=4.3(9)$ from Ref.\,\cite{stone:2005}, where the nuclear magneton is $\mu_\mathrm{N}=\frac{e\hbar}{2m_\mathrm{p}}$. The ZORA transformations of the four component operators were obtained with the implementation \cite{gaul:2024a}. For an explicit discussion of the ZORA expressions for $q_{zz}$ see Ref.\,\cite{gaul:2020b}, for $A_\parallel$ see \cite{gaul:2020} and for the implementation of response functions see Refs.\,\cite{bruck:2023,colombojofre:2022}. A detailed discussion of two-component approximations of the nuclear spin-molecular rotation interaction will be discussed separately. We note that in eq. (\ref{eq:nsmr}) we set $\hat{J}_\mathrm{e}\approx(\bm{r}_i-\bm{r}_\mathrm{CM})\times\hat{\bm{p}}+\frac{\hbar}{2}\bm{\sigma}$, i.e. we omit all small component contributions to the total electronic angular momentum. In contrast to the finite nuclear charge distribution, we assumed a point-like nuclear magnetic dipole moment. 

The employed methodology shows a reasonable agreement with available experimental data with deviations $<1\,\%$ for $B_\mathrm{e}$  and $20\,\%$ for $A_\parallel(\mathrm{F})$ \cite{Ng2022}, and previous four-component relativistic coupled cluster calculations of $A_\parallel(\mathrm{Th})$ \cite{skripnikov:2015} ($7\,\%$ deviation). By computing the difference to a spherical Gaussian magnetization distribution, we estimate the influence of the Bohr--Weisskopf effect to be roughly 3\,\% for $A_\parallel(Th)$. This is irrelevant for the present purpose. Beside uncertainties of roughly 20\,\% in the electronic structure factors, the experimental values of the nuclear magnetic dipole moment ($\gtrsim10\,\%$) and nuclear electric quadrupole moment ($\gtrsim20\,\%$) carry large uncertainties. We also estimated the size of indirect nuclear spin-spin couplings following the approach outlined in Ref.\,\cite{colombojofre:2022} and using the present methodology. We found this effect to be on the order of $600\,\mathrm{Hz}$ and, therefore, negligible for the present discussion.
The relevant molecular parameters are listed in Table \ref{tab:parameter_ThF}.

 The rotational Hamiltonian is given by
  \begin{eqnarray}
  \label{eq:hrot}
    H_{rot} &=& B \mathbf{R}^2=B\left(\mathbf{J}-\mathbf{J_e}\right)^2,
  \end{eqnarray}  
i.e. the nuclear rotational angular momentum is modified my the electronic angular momentum. Since coupling to other electronic states can be neglected, only the diagonal terms of $H_{rot}$ are considered and the rotational eigenstates are given by $E_J^{rot}=B J(J+1)$ \cite{Brown_Carrington}.

The hyperfine interactions are described by Hamiltonian $H_{int}$, Eq.(\ref{eq:hf_hamil}).
The first term in Eq.(\ref{eq:hf_hamil}), $H_Q$, describes the quadrupole interaction, namely the interaction between the quadrupole of the Th-nucleus and the electric field gradient caused by all other charged particles of ThF$^+$ and can be written as \cite{Brown_Carrington}
\begin{eqnarray}
H_Q &=& \frac{eQq_{zz}}{4I_1(2I_1-1)}\left(3I_{1,z}^2-\mathbf{I}_1^2\right) \nonumber \\
&=&\frac{eQq_{zz}}{4I_1(2I_1-1)}\sqrt{6}T^2_{q=0}(\mathbf{I}_1,\mathbf{I}_1),
\label{eq:hf_quad}
\end{eqnarray}
where \changed{the molecule fixed coordinate $z$ is the direction of the internuclear axis}, and spherical tensor notation is used in the second line.
There is no quadrupole interaction for the F-nucleus since its spin is $I_2=1/2$ and only nuclei with spin $I\geq1$ possess a quadrupole moment. The matrix elements of the quadrupole interaction in the $\ket{J\Omega,I_1,F_1,I_2,F,M_F}$ basis are
\begin{eqnarray}
&&\langle J\Omega I_1 F_1 I_2 F M_F|H_Q|J'\Omega I_1 F_1 I_2 F M_F\rangle \nonumber\\
&=&\frac{eQq_{zz}}{4}(-1)^{J+J'-\Omega+I_1+F_1}\begin{Bmatrix}I_1&J'&F_1\\ J&I_1&2 \end{Bmatrix}\begin{pmatrix}J&2&J'\\-\Omega&0&\Omega\end{pmatrix} \nonumber\\
&\times&\sqrt{\frac{(2J+1)(2J'+1)(2I_1+3)(2I_1+2)(2I_1+1)}{2I_1(2I_1-1)}} 
\end{eqnarray}
Thus, the quadrupole interaction allows for transitions between states with $\Delta J = 0,\pm1, \pm2$. Moreover, the quadrupole interaction is independent of $F$ but depends of $F_1$, and thus causes a splitting of the $F_1$-states.

The second and third terms in Eq.(\ref{eq:hf_hamil}) describe the interaction of the nuclear spin with the electronic angular momenta. Considering the projection onto a single electronic state it can be approximated by \cite{Norrgard_2017}
\changed{
\begin{eqnarray}
    H_{IL}+H_{IS} &\approx& A_\Vert (\mbox{Th}) {\bf J}_{e,z} {\bf I}_{1,z}  +A_\Vert (\mbox{F}) {\bf J}_{e,z} {\bf I}_{2,z}  \\
    &=& A_\Vert (\mbox{Th}) \Omega T_0^1(I_1) + A_\Vert (\mbox{F}) \Omega T_0^1(I_2). \nonumber
\label{eq:HIL_HIS}
\end{eqnarray}
where $T_0^1(I)$ the spherical tensor in the molecule fixed frame, and $\Omega$ denotes the projection of the electronic angular momentum onto the internuclear axis ($z$).} 
The non-vanishing matrix elements of the spherical tensor operators
$T_0^1(I_\alpha)$, $\alpha=1,2$ are given by \cite{Norrgard_2017}
\begin{eqnarray}
&&\langle J\Omega I_1 F_1 I_2 F M_F| T_0^1(I_1)|J'\Omega I_1 F_1 I_2 F M_F\rangle \nonumber \\
&=& (-1)^{J'+J+F_1+I_1-\Omega} \begin{Bmatrix}I_1&J'&F_1\\ J&I_1&1 \end{Bmatrix}
\begin{pmatrix}J&1&J'\\-\Omega&0&\Omega\end{pmatrix} \nonumber \\
& \times& \sqrt{(2J+1)(2J'+1)I_1(I_1+1)(2I_1+1)}
\end{eqnarray}
for interaction with the nuclear spin of Th and
\begin{eqnarray}
&& \langle J\Omega I_1 F_1 I_2 F M_F|T_0^1(I_2)|J'\Omega I_1 F_1 I_2 F M_F\rangle \nonumber  \\
&=& (-1)^{2F_1'+F+I_1+I_2-\Omega+1}  
\sqrt{I_2(I_2+1)(2I_2+1)} \nonumber \\ 
&\times&\sqrt{(2J+1)(2J'+1)(2F_1+1)(2F_1'+1)} \nonumber \\
&\times& \begin{Bmatrix}I_2&F_1'&F\\ F_1&I_2&1 \end{Bmatrix} 
\begin{Bmatrix}J'&F_1'&I_1\\ F_1&J&1 \end{Bmatrix}
\begin{pmatrix}J&1&J'\\-\Omega&0&\Omega\end{pmatrix} 
\end{eqnarray}
for the interaction with the nuclear spin of fluorine. 
Both terms depend on $J$ and 
$F_1$, but only the interaction with $I_2$ also depends on $F$ and causes a splitting between the F-states. 
These three terms are the largest contributions to the hyperfine structure of ThF$^+$.

The coupling constants for the remaining contributions are much smaller, cf. table~\ref{tab:parameter_ThF}.
The interaction of the nuclear spin with ${\bf J}$ is given by 
\changed{
\begin{equation}
  H_{IJ} = c(\mbox{Th}) T^1({\bf J}) \cdot T^1({\bf I_1})+
  c(\mbox{F}) T^1({\bf J}) \cdot T^1({\bf I_2}),
  \label{HIJ}
\end{equation}
}
and the matrix elements of the two spherical tensor operators are 
\begin{eqnarray}
&&\langle J\Omega I_1 F_1 I_2 F M_F|T^1(J)\cdot T^1(I_1)|J\Omega I_1 F_1 I_2 F M_F\rangle \nonumber\\
&
=& (-1)^{J+F_1+I_1}\begin{Bmatrix}I_1&J'&F_1\\ J&I_1&1 \end{Bmatrix} \nonumber \\
&\times& \sqrt{J(J+1)(2J+1)I_1(I_1+1)(2I_1+1)} 
\end{eqnarray}
and
\begin{eqnarray}
\label{eq:hJI}
&&\langle J\Omega I_1 F_1 I_2 F M_F|T^1(J)\cdot T^1(I_2)|J\Omega I_1 F_1 I_2 F M_F\rangle \nonumber \\
&
=&(-1)^{J+2F_1'+F+I_1+I_2+1}\begin{Bmatrix}I_2&F_1'&F\\ F_1&I_2&1 \end{Bmatrix}
\begin{Bmatrix}J'&F_1'&I_1\\ F_1&J&1 \end{Bmatrix}  \nonumber \\
&\times& \sqrt{J(J+1)(2J+1)I_2(I_2+1)(2I_2+1)} 
\end{eqnarray}
Again, while both terms depend on $J$ and $F_1$ only the interaction with $I_2$ causes a splitting of the $F$-states.

Note that all matrix elements are diagonal with respect to the orientational quantum number $M_F$. Using Eqs.~(\ref{eq:hrot}) - (\ref{eq:hJI}), the Hamiltonian $H_{mol}$ can be written in the $\ket{J\Omega I_1 F_1 I_2 F M_F}$ basis and its eigenvalues are obtained by numerically diagonalizing the resulting matrix. 
Here, we analyze states with angular momentum $J=1,2$, but we consider states up to $J=8$ to diagonalize $H_{mol}$.
Table~\ref{tab:splitting_components} 
lists the energy differences between states $\ket{J,F_1,F} \longleftrightarrow\ket{J,F_1,F+1}$ and $\ket{J,F_1,F} \longleftrightarrow\ket{J,F_1+1,F}$ for $J=1$ and $J=2$. 

\begin{figure}
   \includegraphics[width=0.9\linewidth]{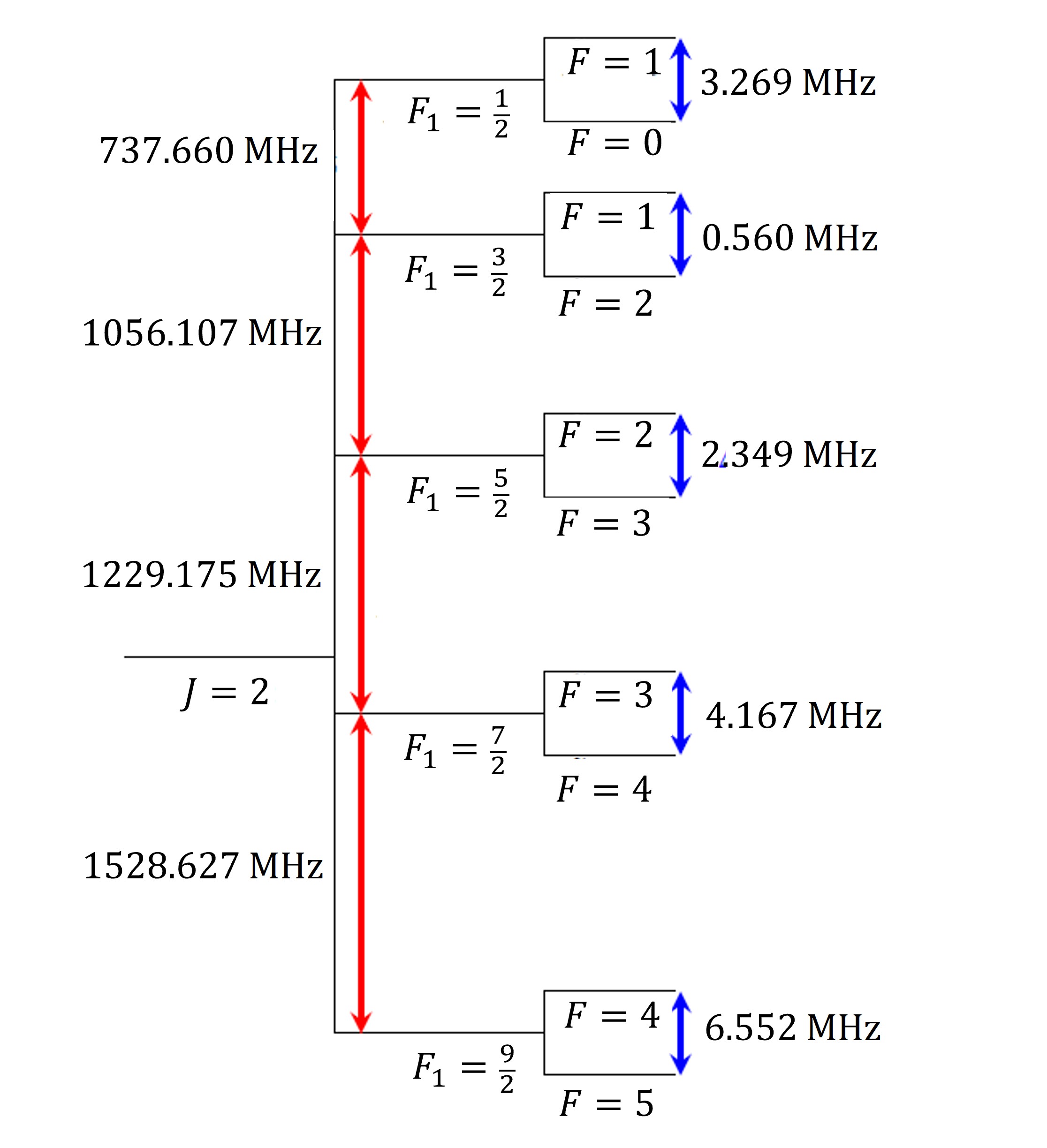}
	\caption{Hyperfine states of $^{229}$Th$^9$F$^+$ in the electronic ground state for $J=2$. }
	\label{fig:spectrum_ThF_J2}
\end{figure}

\changed{
\begin{table*}
\begin{tabular}{|c|c|c|c|c|c|c|}
\hline
$J$&$F_1,F_1'$ & $F,F'$ & $A_\parallel$(Th)$I_{1,z}$&$A_\parallel$(Th)$I_{1,z}$+$A_\parallel$(F)$I_{2,z}$ & $A_\parallel$(Th)$I_{1,z}$+$A_\parallel$(F)$I_{2,z}+H_Q$ & $H_{int}$\\\hline
1&3/2, 3/2 & 1, 2 &      0 &     -5.391 &     -5.770 &     -5.786\\\hline
1&3/2, 5/2 & 2, 2 &  2676.369 &   2676.102 &   2143.245 &   2144.147\\\hline
1&5/2, 5/2 & 2, 3 &    0 &      3.940 &      3.804 &      3.813\\\hline
1&5/2, 7/2 & 3, 3 &   3589.106 &   3581.849 &   4113.993 &   4115.238\\\hline
1&7/2, 7/2 & 3, 4 &    0 &      9.985 &     10.143 &     10.170\\\hline
\end{tabular}
\vspace{0.5cm} \\
\begin{tabular}{|c|c|c|c|c|c|c|}
\hline
$J$ & $F_1,F_1'$ & $F,F'$ & $A_\parallel$(Th)$I_{1,z}$&$A_\parallel$(Th)$I_{1,z}$+$A_\parallel$(F)$I_{2,z}$ & $A_\parallel$(Th)$I_{1,z}$+$A_\parallel$(F)$I_{2,z}+H_Q$ & $H_{int}$\\\hline

2 & 1/2, 1/2 & 0, 1 &     0 &     -3.123 &     -3.241 &     -3.269\\\hline
2 & 1/2, 3/2 & 1, 1 &    450.695 &    451.293 &    738.469 &    739.015\\\hline
2 & 3/2, 3/2 & 1, 2 &     0 &      0.315 &      0.553 &      0.560\\\hline
2 & 3/2, 5/2 & 2, 2 &    816.079 &    814.620 &   1053.762 &   1054.652\\\hline
2 & 5/2, 5/2 & 2, 3 &    0 &      2.300 &      2.325 &      2.349\\\hline
2 & 5/2, 7/2 & 3, 3 &  1247.665 &   1244.288 &   1224.680 &   1225.917\\\hline
2 & 7/2, 7/2 & 3, 4  &  0 &      4.301 &      4.130 &      4.167\\\hline
2 & 7/2, 9/2 & 4, 4  &  1733.771 &   1728.355 &   1521.690 &   1523.267\\\hline
2 & 9/2, 9/2 & 4, 5  &     0 &      6.426 &      6.502 &      6.552\\\hline
\end{tabular}
\caption{Energy splitting $E_{J,F_1,F}-E_{J,F_1',F'}$ of the rotational state $J=1$ and $J=2$ of ThF$^+$ due to individual terms of $H_{int}$, Eq.\ref{eq:hf_hamil}, in MHz.
}
\label{tab:splitting_components}
\end{table*}
}
Note the couplings between different basis states are fairly small (less than 1\%).  
We thus keep the labeling according to the basis states $\ket{J\Omega I_1 F_1 I_2 F}$.
The second column in Table~\ref{tab:splitting_components}  shows the energy splitting induced by the term $A_\parallel(\mbox{Th})I_{1,z}$, with all other contributions switched off. The interaction splits the $F_1$-states in the order of MHz, but does not act on the $F$-states. In the next column, the interaction $A_\parallel(\mbox{F})I_{2,z}$ is added, which accounts for the splitting of the $F$-states of the order of MHz. The quadrupole interaction also contributes considerably to the splitting of the $F_1$ levels, while all remaining terms (last column) only slightly modify the energy levels. 
The energy eigenvalues for the complete Hamiltonian $H_{mol}$ are also depicted in the level diagram Fig.~\ref{fig:spectrum_ThF} for $J=1$ and in Fig.~\ref{fig:spectrum_ThF_J2} for $J=2$.
The hyperfine structure of ThF$^+$ and the resulting splitting of the angular momentum states allows for resonant dipole-phonon coupling of ThF$^+$ ions in a Paul trap, as we discuss in Section~\ref{subsec:resonant_coupling_ThF}. 
\end{appendix}

\bibliography{trap}

\end{document}